\newcommand{\be}{\begin{equation}}
\newcommand{\ee}{\end{equation}}
\newcommand{\bea}{\begin{eqnarray}}
\newcommand{\eea}{\end{eqnarray}}
\newcommand{\nn}{\nonumber}
\newcommand{\p}[1]{(\ref{#1})}
\newcommand\T{\theta_{12}}
\newcommand\Tb{{\bar\theta}_{12}}
\newcommand\Z{Z_{12}}
\newcommand\D{{\cal D}}
\newcommand\Db{\overline{\cal D}}
\documentstyle[12pt]{article}

\begin{document}
\renewcommand{\thefootnote}{\fnsymbol{footnote}}
\thispagestyle{empty}
\begin{flushright}
JINR E2-95-212 \\
LNF-95/026 (P) \\
hep-th/9505142 \\
May 1995
\end{flushright}
\begin{center}
{\large\bf $N=2$ SUPER-$W_3^{(2)}$ ALGEBRA IN SUPERFIELDS}
\vspace{1cm} \\
E. Ivanov${}^{a}\;$\footnote{E-mail: eivanov@thsun1.jinr.dubna.su},
S. Krivonos${}^{a,b}\;$\footnote{E-mail: krivonos@thsun1.jinr.dubna.su} and
A. Sorin${}^{a,b}\;$\footnote{E-mail: sorin@thsun1.jinr.dubna.su}
                     \vspace{1cm}\\
{\it ${}^a$ Bogoliubov Laboratory of Theoretical Physics, JINR,
141980 Dubna, Moscow region, Russia} \\
{\it ${}^b$ INFN-Laboratori Nazionali di Frascati, P.O.Box 13 I-00044 Frascati,
 Italy} \vspace{3cm} \\
{\bf Abstract}
\end{center}
    We present a manifestly $N=2$ supersymmetric formulation of
$N=2$ super-$W_3^{(2)}$ algebra (its classical version) in terms of
the spin 1 unconstrained supercurrent generating a $N=2$ superconformal
subalgebra and the spins 1/2, 2 bosonic and spins 1/2, 2 fermionic
constrained supercurrents. We consider a superfield reduction of
$N=2$ super-$W_3^{(2)}$ to
$N=2$ super-$W_3$ and construct a family of evolution
equations for which
$N=2$ super-$W_3^{(2)}$ provides the second hamiltonian structure.
\vspace{1cm}\\
\begin{center}
{\it Submitted to Physics Letters B}
\end{center}
\setcounter{page}0
\renewcommand{\thefootnote}{\arabic{footnote}}
\setcounter{footnote}0
\newpage

\section{Introduction}

In recent years a plenty of various superextensions of nonlinear
$W$ algebras were constructed and studied from different points of view,
both at
the classical and quantum levels (see, e.g., \cite{s} and references
therein).
An interesting class of bosonic $W$ algebras is so called
quasisuperconformal algebras which include, besides
the bosonic currents with the canonical integer conformal spins, those
with half-integer spins \cite{r,t,f}. The simplest example of such an algebra
is the Polyakov-Bershadsky $W_3^{(2)}$ algebra \cite{p,b}.
It is a bosonic analog of the linear
$N=2$ superconformal algebra (SCA) \cite{x}: apart from two currents with
the spins 2 and 1 (conformal stress-tensor and
$U(1)$ Kac-Moody (KM) current), it contains two bosonic currents with
spins 3/2.
For the currents to form a closed set (with the
relevant Jacobi identities satisfied), the OPE between the spin 3/2
currents should necessarily include a quadratic nonlinearity in the $U(1)$
KM current. So $W_3^{(2)}$, in contrast to its superconformal prototype,
is a nonlinear algebra.

It is natural to seek for supersymmetric extensions of this type of
$W$ algebras and to see how they can be formulated in terms of superfields.
First explicit example of such an extension,
$N=2$ super-$W_3^{(2)}$ algebra, has been constructed
at the classical level in \cite{KS} (its quantum version is
given in \cite{AKS}). It
involves fermionic currents with integer spins 1 and 2
and contains both $N=2$ SCA and $W_3^{(2)}$ as subalgebras.
Actually, it can be regarded as a nonlinear closure of these two algebras.
\footnote{Zamolodchikov's $W_3$ algebra \cite{z}
can be nonlinearly embedded into $W_3^{(2)}$ \cite{lin}, so it also forms a
subalgebra in $N=2$ $W_3^{(2)}$ (in some special basis for the generating
currents of the latter).}

Curiously enough, the spin content of the currents of
$N=2$ super-$W_3^{(2)}$ algebra is such that they cannot be
immediately arranged into $N=2$ supermultiplets with respect to the
$N=2$ SCA which is manifest in the formulation given in \cite{KS}.
This means that $N=2$ super-$W_3^{(2)}$, as it stands, does not admit the
standard $N=2$
superfield description, in contrast, e.g.,
to $N=2$ super-$W_3$ algebra \cite{l,i}. One can still wonder whether
any other superfield formulation exists, perhaps with composite currents
involved into the game. Recall that
it is very advantageous to have a superfield
description because it radically
simplifies computations and allows to present all results in an
explicitly supersymmetric concise form.

In the present paper we show that $N=2$ super-$W_3^{(2)}$ algebra
of ref. \cite{KS} admits a nice superfield description with respect to
{\it another} $N=2$ superconformal subalgebra which is implicit in the
original formulation. An unusual novel feature of this description is that
some of the relevant supercurrents are given by $N=2$ superfields
subjected to nonlinear constraints. Using the superfield formulation
constructed, we demonstrate that $N=2$ super-$W_3$ algebra
follows from $N=2$ super-$W_3^{(2)}$ by a secondary hamiltonian reduction,
like $W_3$ follows from $W_3^{(2)}$ \cite{{delsorb},{lin}}. We also
construct a family of $N=2$ superfield evolution equations with
$N=2$ super-$W_3^{(2)}$ as the second hamiltonian structure.

\section{Preliminaries}

For the reader's convenience we review here the salient features of
$N=2$ super-$W_3^{(2)}$ algebra in terms of component currents \cite{KS}.

A powerful method of constructing conformal
(super)algebras is the hamiltonian reduction method \cite{HR,t,frr}. In
this approach one writes
down a gauge potential $\cal A$ valued in the appropriate
(super)algebra $g$ and then constrain some components of
$\cal A$ to be equal to constants. From the residual gauge
transformations of
the remaining components of $\cal A$ one can immediately read off the
OPEs of some conformal $W$ (super)algebra, with these components
as the generating currents. Since the residual gauge
transformations clearly form a closed set, the Jacobi identities of the
resulting $W$ algebra prove to be automatically satisfied.

A straightforward application of hamiltonian reduction to
superalgebra $sl(3|2)$ gives rise to the classical $N=2$ super-$W_3$
algebra \cite{l}. In \cite{KS} a different choice of constraints
has been made (it corresponds to some non-principal embedding of
$sl(2)$ into the bosonic $sl(3)\times sl(2)$ subalgebra of $sl(3|2)$).
The residual gauge
transformations of the remaining currents yield just
the $N=2$ super-$W_3^{(2)}$ algebra we will deal with here.

More precisely, starting with the following constrained $sl(3|2)$ gauge
potential $\cal A$
\be \label{potential}
{\cal A}=\frac{1}{c} \left(
\begin{array}{ccccc}
2J_s-3J_w & G^{+}     & T_1 & S_1 & S_2 \\
0         & 2J_s-6J_w & G^{-}& 0 & S \\
1 & 0 & 2J_s-3J_w & 0 & S_1 \\
\bar{S}_1 & \bar{S} & \bar{S}_2 & 3J_s-6J_w & T_2 \\
0 & 0 & \bar{S}_1 & 1 & 3J_s-6J_w
\end{array}
\right) ,
\ee
where  $\left\{  J_s,J_w,G^{+},G^{-},T_1,T_2 \right\}$
and $\left\{ S_1,\bar{S}_1,S,\bar{S},S_2,\bar{S}_2 \right\}$
are, respectively, bosonic and fermionic currents, one can easily find
the residual gauge transformations which preserve this particular
form of $\cal A$. They correspond to the following  parameters
\be \label{freepar}
l_1,l_3,a_3,a_5,a_6,a_8,b_3,b_5,(b_1+b_6),c_4,c_5,(c_1+c_6)
\ee
in the standard infinitesimal gauge transformation of $\cal A$
\be \label{var}
\delta {\cal A} = \partial \Lambda + \left[ {\cal A} , \Lambda \right]\;,
\ee
with $\Lambda$ being a $sl(3|2)$-valued matrix of the parameters
\be \label{par}
\Lambda= \left(
\begin{array}{ccccc}
2l_1+l_2+l_3 & a_1     & a_2 & b_1 & b_2 \\
a_3         & 2l_1-2l_3 & a_4& b_3 & b_4 \\
a_5 & a_6 & 2l_1-l_2+l_3 & b_5 & b_6 \\
c_1 & c_2 & c_3 & 3l_1+l_4 & a_7 \\
c_4 & c_5 & c_6 & a_8 & 3l_1-l_4
\end{array}
\right)  .
\ee
The remaining twelve combinations of the parameters are expressed
through these ones and the currents.
After representing these transformations in the form
\bea
\delta\phi (z_1) & = & c \oint dz_2 \left[ -6l_1J_s +18l_3J_w+a_3G^{+}
   +a_6G^{-}+3a_5T_1 - 3a_8T_2+b_3\bar{S}+b_5 \bar{S}_2  \right.\nn \\
 & + & \left. (b_1+b_6)\bar{S}_1 - c_4S_2 - c_5S-(c_1+c_6)S_1
       \right] \phi(z_1)\; , \label{var1}
\eea
where $\phi(z)$ is any current, a self-consistent set of OPEs for
the currents can be extracted from eq.(\ref{var1}).

To understand why this superalgebra was called $N=2$ super-$W_3^{(2)}$,
it is instructive to redefine the currents in the following way
\be
T_w  =  T_1- \frac{1}{c}S_1\bar{S}_1 +\frac{3}{c}J_w^2  \quad ,\quad
T_s  =  -T_2- \frac{1}{c}S_1\bar{S}_1+ \frac{1}{c}J_s^2 .\label{4}
\ee
Then the currents $\left\{  J_w,G^{+},G^{-},T_w \right\}$ and
$\left\{ J_s,S,\bar{S},T_s \right\}$ can be shown to obey
the following  OPEs \footnote{Hereafter, we explicitly write down only
singular terms in OPEs. All the currents appearing in the right hand sides
of the OPEs are evaluated at the point $z_2$ ($z_{12}=z_1-z_2$).}
\bea \label{1}
J_w(z_1)J_w(z_2) & = &  \frac{\frac{c}{6}}{z_{12}^2}  \quad , \quad
J_w(z_1)T_w(z_2)  =  \frac{J_w}{z_{12}^2},  \nn \\
J_w(z_1)G^{\pm}(z_2) & = & \mp \frac{1}{2} \frac{G^{\pm}}{z_{12}} \quad ,\quad
T_w(z_1)G^{\pm}(z_2)  = \frac{3}{2}\frac{G^{\pm}}{z_{12}^2}
+ \frac{{G^{\pm}}'}{z_{12}}, \nn \\
T_w(z_1)T_w(z_2) & = & - \frac{3c}{z_{12}^4}+\frac{2T_w}{z_{12}^2}  +
                         \frac{T_w'}{z_{12}},  \nn \\
G^{+}(z_1) G^{-}(z_2) & = & \frac{2c}{z_{12}^3}-\frac{6J_w}{z_{12}^2}  -
        (T_w-\frac{12}{c}J_w^2+3J_w')\frac{1}{z_{12}},
\eea
\bea\label{2}
J_s(z_1)J_s(z_2) & = & \frac{\frac{c}{2}}{z_{12}^2} \quad , \quad
J_s(z_1)T_s(z_2)  =  \frac{J_s}{z_{12}^2}, \nn \\
J_s(z_1)S(z_2) & = & \frac{1}{2}\frac{S}{z_{12}}  \quad , \quad
J_s(z_1)\bar{S}(z_2)  = -\frac{1}{2}\frac{\bar{S}}{z_{12}}, \nn \\
S(z_1) \bar{S}(z_2) & = & \frac{2c}{z_{12}^3}+\frac{2J_s}{z_{12}^2}  +
        \frac{T_s+J_s'}{z_{12}},  \nn \\
T_s(z_1)S(z_2) & = & \frac{3}{2}\frac{S}{z_{12}^2}+
                         \frac{S'}{z_{12}} \quad , \quad
T_s(z_1)\bar{S}(z_2)  = \frac{3}{2} \frac{\bar{S}}{z_{12}^2}+
                         \frac{{\bar S}'}{z_{12}}, \nn \\
T_s(z_1)T_s(z_2) & = & \frac{3c}{z_{12}^4}+\frac{2T_s}{z_{12}^2} +
                         \frac{T_s'}{z_{12}}\; .
\eea
So they form $W_3^{(2)}$ and $N=2$ SCA with the related central charges.

Thus we are eventually left with the
set of currents which includes those generating $W_3^{(2)}$ and
$N=2$ SCA, as well as four extra fermionic currents
$\left\{ S_1,\bar{S}_1,S_2,\bar{S}_2 \right\}$ with the
integer spins $\left\{ 1,1,2,2 \right\}$. The spin-statistics content
of $N=2$ super-$W_3^{(2)}$ algebra is summarized in Table 1.
\begin{center}
\begin{tabular}{|l|c|c|c|c|c|c|c|c|c|c|c|c|}\hline
 & & & & & & & & & & & &  \\
Currents & $J_s$ & $J_w$ & $S_1$ & ${\bar S}_1$ & $G^{+}$ & $G^{-}$ &
                  $S$   & $\bar S$& $T_s$& $T_w$ & $S_2$& ${\bar S}_2$ \\
                  \hline
 & & & & & & & & & & & &  \\
Spins    & $1^B$ & $ 1^B$ & $ 1^F$ & $1^F$ &$\frac{3}{2}^B$ &$\frac{3}{2}^B$ &
         $\frac{3}{2}^F$ &$\frac{3}{2}^F$ &
          $2^B$ & $ 2^B$ & $ 2^F$ & $2^F$ \\ \hline
\end{tabular} \vspace{0.5cm}\\
{\bf Table 1.}
\end{center}

All the currents with the aforementioned spins, except for
$T_s$ and $T_w$, are primary  with respect to
the following Virasoro stress-tensor $T$ having a zero central charge
\be \label{5}
T=T_s+T_w+\frac{4}{c}S_1\bar{S}_1-\frac{4}{c}J_s^2+\frac{12}{c}J_wJ_s-
  \frac{12}{c}J_w^2  \quad .
\ee
The currents $T_s$ and $T_w$ are  quasiprimary
with the central charges  $3c$ and $-3c$, respectively.
It can be checked that in this $N=2$ super-$W_3^{(2)}$ algebra there
exists no basis for the currents such that all the
currents are primary with respect to some (improved) Virasoro stress-tensor.

The whole set of OPEs of $N=2$ super-$W_3^{(2)}$ algebra
in terms of these currents is given in Appendix.

\setcounter{equation}0
\section{$N=2$ super-$W_3^{(2)}$ algebra in terms of $N=2$ supercurrents}

Despite the fact that $N=2$ super-$W_3^{(2)}$ algebra
has an equal number of bosonic and fermionic currents, it is
unclear how they could be arranged into  $N=2$ supermultiplets.
The main obstruction against the existence of a superfield description is
the fact that in the superalgebra considered the numbers
of currents with integer and half-integer spins do not coincide,
while any $N=2$ superfield clearly contains the equal
number of components with integer and half-integer spins.

To find a way to construct the $N=2$ super-$W_3^{(2)}$ algebra in terms
of $N=2$ superfields, two features of its components OPEs \p{1}, \p{2},
(A.1) must be taken into account.

First of all, $N=2$ super-$W_3^{(2)}$ algebra is nonlinear.
This means that one may choose the basis for its generating currents in
many different ways. The transformations relating different bases
must be invertible but in general they are nonlinear and can include
derivatives of the currents along with the currents themselves.

Secondly, we would like to stress that the OPEs \p{1}, \p{2}, (A.1)
do not fix the
scale of fermionic $S_1, \bar{S}_1, S, \bar{S}, S_2,\bar{S}_2$ and
bosonic $G^{+}, G^{-} $ currents. Moreover, keeping in mind that
all these currents possess definite charges with respect to
the $J_w$ and $J_s$ $U(1)$ currents, one can introduce  a new ``improved''
stress-tensor
\be \label{Timpr}
{\cal T} = T+bJ_w'+gJ_s'\;,
\ee
with respect to which the currents
$G^{+}, G^{-}, S_1, \bar{S}_1, S, \bar{S}, S_2,\bar{S}_2$, still remaining
primary, have the dimensions (spins) listed in Table 2.
\begin{center}
\begin{tabular}{|l|c|c|c|c|}\hline
  & & & & \\
Currents & $G^{+}$ & $G^{-}$ & $S_1$ & ${\bar S}_1$ \\ \hline
  & & & & \\
Spins    & $\frac{3}{2}+\frac{b}{2}+g$ &$ \frac{3}{2}-\frac{b}{2}-g$ &
           $ 1+\frac{b}{6}+\frac{g}{2}$ &$1-\frac{b}{6}-\frac{g}{2}$
                 \\ \hline \hline
  & & & & \\
Currents & $S$ & $\bar S$ & $S_2$ & ${\bar S}_2$ \\ \hline
  & & & & \\
Spins    & $\frac{3}{2}-\frac{b}{3}-\frac{g}{2}$&
    $\frac{3}{2}+\frac{b}{3}+\frac{g}{2}$ &
    $2+\frac{b}{6}+\frac{g}{2}$ & $2-\frac{b}{6}-\frac{g}{2}$
           \\ \hline
\end{tabular} \vspace{0.5cm} \\
{\bf Table 2.}
\end{center}

Thus we cannot exclude the possibility that in some nonlinear basis
the generating currents could have appropriate spins to be
organized into supermultiplets with respect to some new $N=2$ SCA.

Fortunately, just this situation takes place for the superalgebra under
consideration.

In order to demonstrate this, let us pass to the new basis
$(\tilde{J}_s, \tilde{S}_2, \bar{S}_1, \tilde{T}_s)$,
 $(S_1, \tilde{J})$,
 $(G^{+}, \bar{S})$, $ (S, G^{-})$, $(\tilde{T},
\tilde{\bar{S}}_2)$,
related to the original one as
\bea
\tilde{J}_s & = & 4J_s-6J_w  \; , \quad   \tilde{J}=-J_s+3J_w, \nn\\
\tilde{T}_s & = & T-J_s' \; ,  \quad  \tilde{T}=2T_s-2J_s', \nn\\
\tilde{S}_2 & = &S_2+\frac{3}{c}S_1J_s-\frac{3}{c}S_1J_w-S_1',  \nn\\
\tilde{\bar{S}}_2 & = & \bar{S}_2-\frac{1}{c}\bar{S}_1J_s-
  \frac{3}{c}\bar{S}_1J_w+ \bar{S}_1'. \label{newbas}
\eea
All the newly defined currents, except for $\tilde{J}$, are primary with
respect to the Virasoro stress-tensor
$\tilde{T}_s$ (it corresponds to the choice of $b =0$, $g =-1$ in eq.
\p{Timpr} and Table 2) and have the
following spins and statistics: $({1}^B, {3\over 2}^F, {3\over 2}^F, {2}^B)$,
$({1\over 2}^F, {1}^B), ({1\over 2}^B, {1}^F)$, $({2}^F, {5\over 2}^B),
({2}^B, {5\over 2}^F)$.
The spin 1 current $\tilde{J}$ is not even quasiprimary because
its OPE with $\tilde{T}$ contains a central term.

It can be checked that
the currents $(\tilde{J}_s, \tilde{S}_2, \bar{S}_1, \tilde{T}_s)$ form
$N=2$ SCA, while the remaining eight currents ${\tilde J},G^{+},G^{-},S,
{\bar S},S_1,{\tilde{\bar S}}_2,{\tilde T}$
constitute a reducible $N=2$ supermultiplet. Namely,
the sets of the currents $(S_1, \tilde{J})$ and $(G^{+}, \bar{S})$ form
two anti-chiral $N=2$ spin 1/2 supermultiplets, respectively
fermionic and bosonic
ones, with the standard linear transformation properties under
$N=2$ SUSY (the transformation law of fermionic current $S_1$
contains in addition a shift by the transformation parameter).
However, transformation properties of the pairs $(S, G^{-})$
and $(\tilde{T}, \tilde{\bar{S}}_2)$ are more complicated: supersymmetry
$\tilde S$ mixes them with the composite bosonic
and fermionic currents $(B_1, B_2)$ and $(F_1, F_2)$ which have the
spin content $(5/2, 3)$, $(3, 5/2)$ and are defined by
\bea \label{com}
B_1 & = & \frac{1}{c}S_1S \; ,\quad
       B_2=\frac{1}{c}\left( \tilde{J}\tilde{T}
    -G^{+}G^{-}-2S_1\tilde{\bar{S}}_2-S\bar{S} \right), \nn\\
F_1 &=& \frac{1}{c}\left( G^{-}S_1-S\tilde{J}\right) \;, \quad
F_2 =  \frac{2}{c}\left( S_1\tilde{T} -G^{+}S \right) \;.
\eea
These transformation properties follow from the OPEs:
\bea
{\tilde S}_2(z_1)S(z_2) & = & \frac{2B_1}{z_{12}} \; , \nn \\
{\tilde S}_2(z_1)G^{-}(z_2) & = & \frac{2S}{z_{12}^2}+
      \frac{2F_1+S'}{z_{12}} \; , \nn \\
{\tilde S}_2(z_1){\tilde T}(z_2) & = & \frac{F_2}{z_{12}} \; , \nn \\
{\tilde S}_2(z_1){\tilde{\bar S}}_2(z_2) & = & \frac{\tilde T}{z_{12}^2}-
\frac{B_2-\frac{1}{2}{\tilde T}}{z_{12}} \; .
\eea
So the eight currents
$$
(S_1, \tilde{J}),\; (G^{+}, \bar{S}),\; (S, G^{-}), \;
(\tilde{T}, \tilde{\bar{S}}_2)
$$
together form a nonlinear and actually not fully reducible
representation of the $N=2$ SCA defined above.

Crucial for putting this representation in a more transparent manifestly
supersymmetric form is the observation that
the nonlinearly
transforming pairs of
the basic currents, namely $(S, G^{-})$ and $(\tilde{T}, \tilde{\bar{S}}_2)$,
can be combined with the composites $B_1, F_1$ and $F_2, B_2$
into two {\it linearly} transforming spin $2$ $N=2$
supermultiplets with the opposite overall Grassmann parities.

Thus, the basic currents of $N=2$ super-$W_3^{(2)}$ together
with the above composites split into the five irreducible linear
multiplets of the $N=2$ superconformal subalgebra
\bea
\mbox{bosonic, spin 1} & : &
 (\tilde{J}_s, \tilde{S}_2, \bar{S}_1, \tilde{T}_s)\quad  (N=2\; \mbox{SCA})\;
,
        \label{mul1} \\
\mbox{fermionic, spin } \frac{1}{2} & : &
    (S_1, \tilde{J}) \; , \nn \\
\mbox{bosonic, spin } \frac{1}{2} & : &
    (G^{+}, \bar{S}) \; , \label{mul2} \\
\mbox{fermionic, spin 2} & : & (S, G^{-}, B_1, F_1) \; , \nn \\
\mbox{bosonic, spin 2} & : &
(\tilde{T}, \tilde{\bar{S}}_2, F_2, B_2)\;. \label{mul3}
\eea

This extended set of currents, in accordance with their
spin content, is naturally accomodated by the five $N=2$
supercurrents:
general spin 1 $J(Z)$, spin 1/2 anti-chiral fermionic $G(Z)$ and bosonic
$Q(Z)$, spin 2 fermionic $F(Z)$ and bosonic $T(Z)$
\footnote{By $Z$ we denote the coordinates of $1D$ $N=2$ superspace,
$Z=(z, \theta, \bar{\theta})$.}. The precise relation of the
components of these superfields to the currents of $N=2$ $W_3^{(2)}$
is quoted in Appendix. Below we reformulate $N=2$ $W_3^{(2)}$ in terms of
SOPEs of these supercurrents.

The superfield $J(Z)$ generates the $N=2$ SCA with SOPE
\begin{equation} \label{Jope}
J(Z_1)J(Z_2)=\frac{-2c+\T\Tb J}{\Z^2}+ \frac{\Tb\Db J- \T\D J
 +\T\Tb J'}{\Z} \; ,
\end{equation}
where
\begin{equation}
\T=\theta_1-\theta_2 \quad , \quad \Tb=\bar\theta_1-\bar\theta_2 \quad ,
 \quad \Z=z_1-z_2+\frac{1}{2}\left( \theta_1\bar\theta_2
-\theta_2\bar\theta_1 \right) \quad ,
\end{equation}
and $\D,\Db$ are the spinor covariant derivatives defined by
\begin{equation}
\D=\frac{\partial}{\partial\theta}
 -\frac{1}{2}\bar\theta\frac{\partial}{\partial z} \quad , \quad
\Db=\frac{\partial}{\partial\bar\theta}
 -\frac{1}{2}\theta\frac{\partial}{\partial z} \; ,
\end{equation}
$$
\left\{\D,\Db \right\}= -\frac{\partial}{\partial z} \quad , \quad
\left\{\D,\D \right\} = \left\{\Db,\Db \right\}= 0.
$$
The next SOPEs express the property that the remaining four
supecurrents have
the aforementioned spins with respect to this $N=2$ SCA
\begin{equation}\label{Gprim}
J(Z_1)G(Z_2) = \frac{-c\T +\frac{1}{2}\T\Tb G}{\Z^2}+
 \frac{\T\Tb  G' -\T\D G - G}{\Z} \quad ,
\end{equation}
\begin{eqnarray}
J(Z_1)Q(Z_2) & = & \frac{\frac{1}{2}\T\Tb Q}{\Z^2}+
 \frac{\T\Tb  Q' -\T\D Q - Q}{\Z} \quad , \nonumber \\
J(Z_1)F(Z_2) & = & \frac{2\T\Tb F}{\Z^2}+
 \frac{\T\Tb  F' + \Tb \Db F -\T\D F }{\Z} \quad , \nonumber \\
J(Z_1)T(Z_2) & = & \frac{2\T\Tb T}{\Z^2}+
 \frac{\T\Tb  T' + \Tb \Db T -\T\D T }{\Z} \quad . \label{QFTprim}
\end{eqnarray}
Let us pay attention to the presence of a central term in \p{Gprim}. It
reflects the property that
the superfield $G(Z)$ transforms inhomogeneously under $N=2$ SCA. All other
superfields are primary with respect to the $N=2$ SCA supercurrent $J(Z)$.

In each of the supercurrents $F(Z)$ and $T(Z)$, the spin 3 component
and one of the spin $5/2$ components are composite (see \p{com}).
In the superfield language, this implies that
these superfields
have to satisfy some constraints. Using the formulas (A.2) of Appendix,
one can check that the relations \p{com} amount to the following nonlinear
constraints
\begin{eqnarray}
A_1 & \equiv & \Db F +\frac{2}{c} (G F) =0 \quad , \label{con1} \\
A_2 & \equiv & \Db T +\frac{2}{c} (G T) +\frac{2}{c} (QF) =0 \quad .
      \label{con2}
\end{eqnarray}
For completeness, we add also the chirality conditions for $G$, $Q$
\be  \label{conchir}
\Db G = 0\;, \;\;\; \;\;\Db Q =0\;.
\ee

By means of eq. \p{con2} one could, in principle, eliminate $F(Z)$ in
terms of  $T(Z), G(Z)$ and $Q(Z)$.
If one substitutes this expression for $F(Z)$ in the
constraint \p{con1}, the latter is satisfied identically.
However, this expression is
singular at $Q(Z)=0$. We
prefer to deal with two constrained supercurrents in order to
have polynomial non-singular expressions in all SOPEs.

Now we are ready to construct the remaining SOPEs of $N=2$ super-$W_3^{(2)}$.
Taking the most general Ansatz for these SOPEs in terms of the introduced
superfields, using \p{con1}, \p{con2}, \p{com} and requiring the latter
to be
consistent with the  OPEs for the superfield
components (see Appendix) we obtain the following non-trivial relations
\begin{eqnarray}
G(Z_1)Q(Z_2) & = & -\frac{\frac{1}{2} \T Q}{\Z} ,\nonumber \\
G(Z_1)F(Z_2) & = & \frac{\frac{1}{2} \T F}{\Z} ,\nonumber \\
G(Z_1)T(Z_2) & = & -2\frac{c\T + \T\Tb G}{\Z^3}
   - \frac{\T\Tb (  G' -\frac{2}{c} G\D G-
  \frac{1}{c}JG-\frac{1}{2}\Db J)}{\Z^2} \nonumber \\
 & & +\frac{\T ( J +2\D G) -2G}{\Z^2}
+\frac{\Db J - 2 G' +\frac{4}{c} G\D G+\frac{2}{c} JG}{\Z} \nonumber \\
Q(Z_1)F(Z_2) & = & 2\frac{c\T + \T\Tb G}{\Z^3}+
  \frac{2G-\T (J +2\D G)+\T\Tb ( G' -\frac{2}{c} G\D G-
  \frac{1}{c}JG-\frac{1}{2}\Db J)}{\Z^2} \nonumber \\
 & & +\frac{2 G' -\frac{4}{c} G\D G- \frac{2}{c} JG- \Db J +
    \frac{1}{2}\T T}{\Z}\; , \nonumber \\
Q(Z_1)T(Z_2) & = & -2\frac{\T\Tb Q}{\Z^3}+
  \frac{2\T\D Q-2Q-\T\Tb (Q' -\frac{2}{c} \D G Q -
  \frac{1}{c}JQ-\frac{2}{c} G\D Q) }{\Z^2} \nonumber \\
 & & +\frac{ \frac{4}{c} \D G Q+\frac{4}{c} G\D Q + \frac{2}{c} JQ-
        2 Q'}{\Z} \; , \nonumber \\
F(Z_1)T(Z_2) & = & \frac{\T\Tb(\frac{2}{c} JF -3 F'+
     + \frac{12}{c}\D G F)+
        2\T \D F+ 2\Tb \Db F+4F }{\Z^2} \nonumber \\
& & +  \frac{\T\Tb (-\frac{1}{2} F'' -
  \frac{3}{2} [\D, \Db ] F'+
  \frac{2}{c} J F'-\frac{3}{c}\D J \Db F -\frac{2}{c}\Db J \D F+
  \frac{2}{c} \D G' F)}{\Z} \nonumber \\
& & + \frac{ \T\Tb ( \frac{2}{c} \D G F'+
  \frac{6}{c} G' \D F+\frac{6}{c} G \D F'-
  \frac{2}{c^2} \D J GF) }{\Z} \nonumber \\
& & +\frac{\T( 2 \D F' + \frac{2}{c} \D JF -\frac{2}{c} J \D F-
 \frac{4}{c} \D F \D G) -
 \Tb( \frac{2}{c} \Db J F -\frac{1}{c} J\Db F -
        \frac{3}{c} \Db F') }{\Z}\nonumber \\
& & +\frac{\Tb(\frac{2}{c} G' F+ \frac{6}{c} G F'
+\frac{3}{c}\D G \Db F-\frac{2}{c^2} JGF-\frac{2}{c^2}(\D G GF)) +
         2 F'}{\Z} \; , \nonumber \\
T(Z_1)T(Z_2) & = & \frac{4 T +2\T \D T +2\Tb \Db T +
  \T\Tb(\frac{2}{c} JT-\frac{12}{c} F \D Q +\frac{12}{c}\D G T-
  3 T')}{\Z^2} \nonumber \\
& & +\frac{\Tb( 2\Db T'
   +\frac{4}{c}Q F'+\frac{8}{c^2}GF\D Q+\frac{4}{c}G T'
  +\frac{2}{c} J\Db T+\frac{4}{c}\D G\Db T-\frac{2}{c} T\Db J)}{\Z}
\nonumber \\
& & +\frac{2 T' +\T(2 \D T'+
 \frac{4}{c}\D F\D Q -\frac{2}{c}J\D T-\frac{4}{c}\D G \D T+
  \frac{2}{c} T\D J)}{\Z} \nonumber \\
& & -\frac{ \T\Tb(  T''
+ [\D,\Db ] T' -\frac{2}{c} \D T\Db J+\frac{4}{c} \D T  G'
+\frac{4}{c} \D F Q' -\frac{8}{c^2} G\D F\D Q)}{\Z} \nonumber \\
& & -\frac{\T\Tb( \frac{8}{c^2} F\D G\D Q-\frac{2}{c}J T'+
   \frac{4}{c}\D J\Db T
+\frac{2}{c}[\D,\Db ]F\D Q+\frac{4}{c} \D T' G)}{\Z} \nonumber \\
& & -\frac{\T\Tb( \frac{4}{c}\D F' Q+
\frac{2}{c} F' \D Q -\frac{4}{c^2} QF\D J-
\frac{4}{c^2}GT\D J)}{\Z}\nonumber \\
& & -\frac{\T\Tb(\frac{4}{c} F\D Q' -\frac{4}{c} \D G T'-
\frac{4}{c} \D G' T)}{\Z} \; . \label{OPE}
\end{eqnarray}

The above SOPEs are self-consistent only on the shell of
constraints \p{con1}, \p{con2}. These constraints
are first class and the Jacobi identities are satisfied only on
their shell $A_1 = A_2 = 0$. They are consistent with the SOPEs
\p{Jope}, \p{Gprim}, \p{QFTprim}, \p{OPE} in the sense that the
SOPEs of $A_1$, $A_2$ with all supercurrents
are vanishing on the constraint shell (the compatibility of the
whole set of SOPEs with the linear chirality conditions \p{conchir}
is evident by construction).
It should be pointed out that it is impossible to satisfy
the Jacobi identities off the constraint shell
unless
one further enlarges the set of supercurrents. We have checked this
by inserting the expressions
$A_1$ and $A_2$ \p{con1}, \p{con2}
\footnote{$A_1$ and $A_2$ are non-zero off the constraints
shell.}
in all appropriate places in the right hand sides of the SOPEs obtained.
Thus the constraints \p{con1}, \p{con2} are absolutely necessary for
the above set of $N=2$ superfields to form a closed algebra. In a
forthcoming paper devoted to $N=2$ superfield hamiltonian
reduction \cite{AIS} it will be shown that these
constraints (as well as the chirality conditions \p{conchir})
are remnants of the Hull-Spence type constraints \cite{HS}
for the supercurrents of $N=2$ extension of affine superalgebra
$sl(3|2)^{(1)}$.

Our final remark concerns the presence of the spin $1/2$ currents
$S_1$ and $G^+$ in the basis \p{newbas}. At first sight, following the
reasonings of ref. \cite{godschw}, one could think that they can be
factored out to yield a smaller nonlinear algebra. However, this is not
true in the present
case because an important assumption of ref. \cite{godschw} does not
hold, namely the assumption that OPEs
between the spin 1/2 currents contain singularities. Indeed,
the OPEs of these currents are regular in our case. So
in the algebra $N=2$ super-$W_3^{(2)}$ the spin 1/2
currents cannot be removed.

\setcounter{equation}0
\section{Superfield reduction to $N=2$ super-$W_3$ algebra}

In this section we show that, if one imposes the additional
first-class constraint on the supercurrent $Q(Z)$
\be \label{con3}
\tilde{Q}(Z) \equiv Q(Z)-c=0
\ee
and the condition
\be  \label{con4}
G(Z)=0
\ee
which fixes a gauge with respect to gauge transformations generated by
\p{con3} (together \p{con3} and \p{con4} form a set of
second-class constraints), one arrives at SOPEs of some self-consistent
nonlinear algebra written
in terms of unconstrained supercurrents. This algebra turns out to be
none other than the well-known $N=2$ super-$W_3$
algebra \cite{l} formulated in terms of $N=2$ superfields in \cite{i}.

Let us define new superfields $\tilde{J}(Z)$ and $\tilde{T}(Z)$
\begin{eqnarray}
\tilde{J}(Z) & = & J(Z)-2\D G(Z) +2\partial Q(Z), \nonumber \\
\tilde{T}(Z) & = & T(Z)-\frac{1}{3}[\D,\Db ]J(Z)+\frac{4}{3}\D \partial G(Z)-
\frac{4}{9c}J(Z)^2 - \frac{20}{9c}(\partial J(Z)Q(Z))+ \nonumber \\
& & \frac{2}{9c}J(Z)\partial Q(Z) -
  \frac{2}{c}\Db J(Z)\D Q(Z)-
\frac{16}{9c}\D G(Z)\D G(Z) +\frac{20}{9c}\partial Q(Z) \partial Q(Z)+ \nn \\
  & &\partial J(Z) - \frac{2}{3}\partial^2 Q(Z)\; .
\end{eqnarray}
This substitution is dictated by the requirement that SOPEs of these
supercurrents with $G(Z)$
and $\tilde{Q}(Z)$ \p{con3} be homogeneous in $\tilde{Q}(Z)$ and $G(Z)$.
The supercurrent
$\tilde{J}(Z)$ can be checked to generate another $N=2$ SCA, such
that the conformal weights of $\tilde{Q}(Z), G(Z), T(Z)$ and $F(Z)$ with
respect to it equal $0, 1/2, 2$ and
$5/2$, respectively. The constraints \p{con3} and \p{con4} prove to be
preserved by this $N=2$ SCA.

Thus the superfields $\tilde{J}(Z)$ and $\tilde{T}(Z)$ by construction
are gauge invariant with respect to the gauge transformations generated by
the first-class constraints \p{con3}. So, according to the standard ideology
of hamiltonian reduction \cite{HR,t,frr} \footnote{Actually, the described
procedure supplies a nice example of secondary hamiltonian reduction in
$N=2$ superspace \cite{AIS}.}, they
have to form a closed superalgebra (with all the Jacobi identities
satisfied) on the shell of constraints \p{con3}, \p{con4} and with
\begin{equation}\label{216}
F=-\frac{1}{2}\Db T.
\end{equation}
The last relation follows by substituting \p{con3}, \p{con4} into
eq.\p{con2}. Note that with this $F$ eq.\p{con1} is identically
satisfied.

Using the SOPEs of $N=2$ super-$W_3^{(2)}$
we find  that the resulting
SOPEs for the currents $\tilde{J}(Z)$ and $\tilde{T}(Z)$
after substituting \p{con3}, \p{con4}, \p{216}
exactly coincide with
SOPEs of classical $N=2$ super-$W_3$
algebra \cite{i}. In the next Section we will make use of this result to
construct the simplest nontrivial hamiltonian flow on
$N=2$ super-$W_3^{(2)}$.

\setcounter{equation}0
\section{Generalized $N=2$ super Boussinesq equation}

The most general hamiltonian which can be constructed out of the
five superfields
$\tilde{J}(Z), G(Z)$, $\tilde{Q}(Z), \tilde{T}(Z)$ and $F(Z)$ of
$N=2$ super-$W_3^{(2)}$ algebra under the natural assumptions that it (i)
respects rigid $N=2$ supersymmetry and (ii) has the same scaling
dimension 2 as the hamiltonian of the ordinary bosonic
Boussinesq equation, is given by
\be
{\cal H} =\int dzd\theta d{\bar\theta} \left(
T+v_1J^2 +v_2 J\D G +v_3 J\partial Q+v_4\D G \partial Q \right)\;.
           \label{ham}
\ee
Note the presence of the free parameters $v_1, ... , v_4$ in \p{ham}.
Now, using SOPEs of $N=2$ super-$W_3^{(2)}$ algebra and the definition
\be
\frac{\partial \phi}{\partial t} = \left\{ \phi , {\cal H}\right\}
            \label{eqm}
\ee
(here $\phi(z)$ is any supercurrent of $N=2$ super-$W_3^{(2)}$ and
the Poisson brackets in the r.h.s. of \p{eqm} are understood),
it is straightforward to find the explicit form of the evolution equations.
Due to the complexity of these equations, it is not so
illuminating to write down them here. We also postpone
to future publications the analysis of integrability of this
system.

In ref. \cite{i} we have constructed,
in $N=2$ superfield form, the most
general one-parameter super Boussinesq equation with the second
hamiltonian structure given by the classical $N=2$ super-$W_3$ algebra.
With making use of the results of Sect. 4 it is not difficult to
show that the
obtained system of evolution equations reproduces the one of ref.
\cite{i} upon the above truncation
of $N=2$ super-$W_3^{(2)}$ to $N=2$ super-$W_3$ and with the following
relations between the parameters in \p{ham}
\be
v_1=\frac{4\alpha-\frac{8}{3c}}{6},\quad v_2=-\frac{4}{c}-4v_1\,
\quad v_3=4v_1 \; ,
\ee
Here $\alpha $ is the parameter entering the $N=2$ super Boussinesq
equation \cite{i}.

\section{Conclusion}

To summarize, we have concisely rewritten classical $N=2$ super-$W_3^{(2)}$
algebra of ref. \cite{KS} in terms of five constrained $N=2$ superfields,
found its
superfield reduction to $N=2$ super-$W_3$ algebra \cite{l,i}, and
constructed
a family of $N=2$ supersymmetric equations the second hamiltonian
structure for which is given by this superalgebra and which generalize the
$N=2$ super Boussinesq equation of ref. \cite{i}. In a
forthcoming publication \cite{ahn} we will extend our consideration to the
case of full quantum $N=2$ super-$W_3^{(2)}$ algebra.

An interesting
problem is to find out possible string theory implications of
$N=2$ $W_3^{(2)}$ algebra, both in its component and superfield formulations.
The fact that there exists a zero central charge stress-tensor (\ref{5})
with respect to which almost all of the currents are primary suggests
that this algebra admits an
interpretation as a kind of twisted topological superconformal algebra
and so has a natural realization in terms of BRST structure associated
with some string (the $W_3^{(2)}$ one?) or superstring.

\section*{Acknowledgments}
We would like to thank C. Ahn, S. Bellucci and V. Ogievetsky for
many useful and clarifying discussions.
Two of us (S.K. and A.S.) thank Laboratori Nazionali di Frascati for the
hospitality during the course of this work.

We acknowledge a partial
support from the Russian Foundation of Fundamental Research,
grant 93-02-03821, and the International Science Foundation, grant M9T000.

\section*{Appendix}
Here we present the component OPEs for the $N=2$ super-$W_3^{(2)}$
algebra \cite{KS} and give the relation between the currents of the latter
and the components of $N=2$ supercurrents $J(Z)$, $G(Z)$, $Q(Z)$, $F(Z)$ and
$T(Z)$.

The whole set of the OPEs contains, besides those of
the subalgebras $W_3^{(2)}$ and $N=2$ SCA \p{1}, \p{2}, the following
non-trivial relations:
\bea
& & J_w(z_1)S_1(z_2)  =  -\frac{\frac{1}{6}S_1}{z_{12}} , \quad
J_s(z_1)S_1(z_2)  =  -\frac{\frac{1}{2}S_1}{z_{12}} ,\;
J_s(z_1)J_w(z_2)  =  \frac{\frac{c}{3}}{z_{12}^2} , \nn \\
& & J_s(z_1)T_w(z_2)  =  \frac{2J_w}{z_{12}^2} \quad, \quad
J_s(z_1)G^{+}(z_2)  =  -\frac{G^{+}}{z_{12}}  , \;
J_w(z_1)S(z_2)  =  \frac{\frac{1}{3}S}{z_{12}} , \nn \\
& & J_s(z_1)S_2(z_2)  =  -\frac{\frac{1}{2}S_2}{z_{12}} \quad, \quad
J_w(z_1)T_s(z_2)  =  \frac{\frac{2}{3}J_s}{z_{12}^2} \quad, \quad
J_w(z_1)S_2(z_2)  =  -\frac{\frac{1}{6}S_2}{z_{12}} , \nn \\
& & T_s(z_1)T_w(z_2)  =  \frac{
 \frac{4}{c}\left( S_1\bar{S}_1+J_wJ_s \right)}{z_{12}^2}
                +\frac{
 \frac{2}{c}\left( S_1\bar{S}_2+S_1\bar{S}'_1-S_2\bar{S}_1+
                S_1'\bar{S}_1+2J_wJ_s'\right)}{z_{12}} , \nn \\
& & T_s(z_1)G^{+}(z_2)  =  -\frac{
  \frac{2}{c}\left( G^{+}J_s-S_1\bar{S}\right)}{z_{12}}, \quad
T_s(z_1)S_1(z_2)  =  -\frac{\frac{1}{2}S_1}{z_{12}^2}+
           \frac{S_2-S_1'-
 \frac{1}{c}\left( S_1J_s+3J_wS_1\right)}{2z_{12}} , \nn \\
& & T_s(z_1)S_2(z_2)  =  -\frac{3S_1}{z_{12}^3}+
           \frac{2S_2-3S_1' +\frac{1}{c}\left(
3S_1J_s-9J_wS_1\right)}{2z_{12}^2}
          +  \frac{\frac{2}{c}G^{+}S-\frac{4}{c}S_1T_s+
       \frac{3}{c^2}S_1J_s^2+\frac{1}{c}S_1J_s'}{2z_{12}}  \nn \\
& &       -   \frac{\frac{3}{c}S_2J_s-\frac{3}{c}J_wS_2-\frac{6}{c^2}J_wJ_sS_1+
          \frac{9}{c^2}J_w^2S_1+\frac{6}{c}J_wS_1'-\frac{2}{c}S_1'J_s+
           \frac{3}{c}J_w'S_1-S_2'+S_1''}{2z_{12}} , \nn \\
& & T_w(z_1)S_1(z_2)  =  -\frac{\frac{1}{2}S_1}{z_{12}^2}-
           \frac{S_2-\frac{1}{c}S_1J_s+\frac{5}{c}J_wS_1+S_1'}{2z_{12}}
 \quad, \quad
T_w(z_1)S(z_2)  =
           -\frac{\frac{2}{c}G^{-}S_1-\frac{2}{c}J_wS}{z_{12}} , \nn \\
& & T_w(z_1)S_2(z_2)  =  \frac{3S_1}{z_{12}^3}+
           \frac{2S_2-\frac{3}{c}S_1J_s+\frac{9}{c}J_wS_1+3S_1'}{2z_{12}^2} -
        \frac{\frac{2}{c}G^{+}S+\frac{4}{c}S_1T_w -\frac{1}{c^2}S_1J_s^2+
        \frac{1}{c}S_1J_s'}{2z_{12}}  \nn \\
& &    -  \frac{ \frac{1}{c}S_2J_s-\frac{1}{c}J_wS_2+\frac{6}{c^2}J_wJ_sS_1-
          \frac{21}{c^2}J_w^2S_1-\frac{6}{c}J_wS_1'+\frac{2}{c}S_1'J_s
          -\frac{3}{c}J_w'S_1-S_2'-S_1''}{2z_{12}} , \nn \\
& & G^{+}(z_1)S(z_2)  =  -\frac{2S_1}{z_{12}^2}-
           \frac{S_2-\frac{1}{c}S_1J_s-\frac{3}{c}J_wS_1+S_1'}{z_{12}}
  \quad, \quad
G^{+}(z_1)S_2(z_2)  = -
           \frac{\frac{3}{2c}G^{+}S_1}{z_{12}}  , \nn \\
& & G^{-}(z_1)S_1(z_2)  = -
           \frac{\frac{1}{2}S}{z_{12}} \quad, \quad
G^{-}(z_1)S_2(z_2)  =  \frac{\frac{3}{2}S}{z_{12}^2}-
\frac{\frac{1}{c}G^{-}S_1+\frac{1}{c}J_sS-\frac{9}{c}J_wS-S'}{2z_{12}}
   , \nn \\
& & S_1(z_1)\bar{S}(z_2)  =
           \frac{G^{+}}{2z_{12}} \quad, \quad
S_1(z_1)\bar{S}_2(z_2)  =
           \frac{T_s+T_w+\frac{2}{c}S_1\bar{S}_1-\frac{1}{c}J_s^2-
            \frac{3}{c}J_w^2}{2z_{12}} , \nn \\
& & S(z_1)S_2(z_2)  =
           \frac{3S_1S}{2cz_{12}} \quad, \quad
S(z_1)\bar{S}_2(z_2)  =  -\frac{3G^{-}}{2z_{12}^2}-
           \frac{\frac{3}{c}G^{-}J_s+\frac{1}{c}\bar{S}_1 S-
           \frac{3}{c}J_wG^{-}+{G^{-}}'}{2z_{12}} , \nn \\
& & S_2(z_1)S_2(z_2)  =
           \frac{2S_1S_2}{cz_{12}} , \;
  S_1(z_1)\bar{S}_1(z_2)  =  -\frac{\frac{c}{2}}{z_{12}^2}+
          \frac{\frac{3}{2}J_w-\frac{1}{2}J_s }{z_{12}}  \nn \\
& & S_2(z_1)\bar{S}_2(z_2)  =  \frac{3c}{z_{12}^4}+
     \frac{3J_s-9J_w}{z_{12}^3}
    +  \frac{2T_s-2T_w+\frac{1}{c}J_s^2-\frac{18}{c}J_wJ_s+
          \frac{33}{c}J_w^2+3J_s'- 9J_w'}{2z_{12}^2} \nn
\eea
$$
 + \frac{\frac{1}{c}G^{+}G^{-}+\frac{1}{c}S_1\bar{S}_2+
   \frac{1}{c}S\bar{S}+\frac{1}{c}S_2\bar{S}_1
   -\frac{1}{2c^2}J_s^3
  +  \frac{1}{c}T_sJ_s-\frac{1}{c}T_wJ_s-
   \frac{3}{c}J_wT_s+\frac{3}{c}J_wT_w -\frac{3}{2c^2}J_wJ_s^2}{z_{12}}
$$
$$
 +  \frac{\frac{33}{2c^2}J_w^2J_s-\frac{45}{2c^2}J_w^3-
   \frac{9}{2c}J_wJ_s'+\frac{1}{2c}J_s'J_s -\frac{9}{2c}J_w'J_s
 +   \frac{33}{2c}J_w'J_w+\frac{1}{2}T_s'-\frac{1}{2}T_w'
          +\frac{1}{2}J_s''-\frac{3}{2}J_w''}{z_{12}} . \eqno(A.1)
$$
Here we omitted the OPEs which can be obtained from (A.1) via
the discrete automorphisms:  $J_{w,s}\rightarrow -J_{w,s}$,
$G^{\pm}\rightarrow \pm G^{\mp}$, $S\rightarrow {\bar S}$,
${\bar S}\rightarrow S$, $S_1\rightarrow {\bar S}_1$,
${\bar S}_1\rightarrow -S_1$,$S_2\rightarrow -{\bar S}_2$,
${\bar S}_2\rightarrow  S_2$.

The currents are related to the components of the $N=2$ $W_3^{(2)}$
supercurrents in the following way
$$
J| =  {\tilde J}_s ,\; \Db J|=\sqrt{2}{\tilde S}_2 ,\;
      \D J|=\sqrt{2}{\bar S}_1 ,\;
         \left[\D ,\Db\right] J|=-2{\tilde T}_s ,
$$
$$
G| =  -\sqrt{2}S_1 ,\; \D G| = {\tilde J} , \;
Q| =  \sqrt{2}G^{+} , \; \D Q| = {\bar S} ,
$$
$$
F| =  S , \; \D F| = -\frac{1}{\sqrt{2}}G^{-} , \;
           \Db F =2\sqrt{2}B_1 , \; \D\Db F|=2F_1 ,
$$
$$
T| =  {\tilde T} , \; \D T| = -\sqrt{2}{\tilde{\bar S}}_2 , \;
           \Db T|=\sqrt{2}F_2 ,\; \D\Db T| =-2B_2 . \eqno(A.2)
$$
The composite currents $B_1, B_2, F_1, F_2$ were defined in \p{com}.

\end{document}